\documentclass[onecolumn,showpacs,preprintnumbers,amsmath,amssymb,preprint]{revtex4}
\usepackage{amssymb}
\usepackage[dvips]{graphicx}
\begin{document}

\title{Normal state  diamagnetism of  charged bosons in cuprate superconductors}
\author{A. S. Alexandrov}

\affiliation{Department of Physics, Loughborough University, Loughborough LE11 3TU, United Kingdom\\
}

\begin{abstract}
 Normal state orbital  diamagnetism of charged
bosons quantitatively accounts for recent high-resolution
magnetometery results near and above the resistive critical
temperature $T_c$ of superconducting cuprates. Our parameter-free
descriptions of normal state diamagnetism, $T_c$, upper critical
fields and specific heat anomalies unambiguously support the 3D
Bose-Einstein condensation of preformed real-space pairs  with zero
off-diagonal order parameter above $T_c$ at variance with phase
fluctuation scenarios of cuprates.

\end{abstract}

 \maketitle

A possibility of real-space pairing, as opposed to the Cooper
pairing, has been the subject of many discussions,  particularly
heated over the last 20 years after the discovery of high
temperature superconductivity in cuprates \cite{bed}. The first
proposal for high temperature superconductivity, made by Ogg Jr in
1946 \cite{ogg}, already involved  real-space pairing of individual
electrons into bosonic molecules  with zero total spin. This idea
was further developed as a natural explanation of conventional
superconductivity by Schafroth  and Butler and Blatt \cite{sha}.
However, with one or two exceptions, the Ogg-Schafroth picture was
condemned and practically forgotten because it neither accounted
quantitatively for the critical behavior of conventional (i.e. low
$T_{c}$) superconductors, nor did it explain the microscopic nature
of attractive forces which could overcome the Coulomb repulsion
between two electrons  constituting a
 pair. The failure of the `bosonic' picture of individual
electron pairs became fully transparent when Bardeen, Cooper and
Schrieffer \cite{bcs} proposed that two electrons in a
superconductor were indeed  correlated, but on a very large distance
of about $10^{3}$ times of the average inter-electron spacing.

Highly successful for low-Tc metals and alloys  the BCS theory has
led many researchers to believe that novel high-temperature
superconductors should also be  "BCS-like". However, the
Ogg-Schafroth and the BCS descriptions are actually two opposite
extremes of the same electron-phonon interaction. Indeed by
extending the BCS theory towards the strong interaction between
electrons and ion vibrations, a charged Bose gas (CBG) of tightly
bound electron pairs surrounded by lattice deformations (i.e. of
{\it small bipolarons}) was predicted by us \cite{aleran} with a
further prediction  that high $T_c$ should exist in the crossover
region of the electron-lattice interaction strength from the
BCS-like to bipolaronic superconductivity \cite{ale}. Experimental
evidence for an exceptionally strong electron-phonon interaction in
novel superconductors \cite{zhao,LAN,mic,ega} is
 so overwhelming that bipolaronic CBG \cite{alebook} could be a feasible
alternative to  BCS-like scenarios of cuprates.  Nevertheless,
some authors \cite{kiv} have dismissed any real-space pairing, advocating  a \textit{%
collective} pairing (i.e  \textit{\ Cooper }pairs in the momentum
space) at some high temperature $T^{\ast}$ which are  phase coherent
below a lower temperature  $T_c < T^{\ast }$. Ref.\cite{kiv} has
argued that the superconducting  transition in cuprates is an almost
two-dimensional Kosterlitz-Thouless (KT) transition, where a vortex
liquid exists above $T_c$ different from  the BCS theory and  its
strong-coupling bipolaronic extension with a perfectly "normal"
state without any off-diagonal order.

So far there has been no decisive conclusion  on the origin of
anomalous normal state  of cuprates.  Some normal state properties
have been satisfactorily interpreted within the Fermi-liquid
approach, while many others have been understood with preformed
real-space \cite{alebook} or Cooper \cite{kiv} pairs, in particular
on the underdoped side of the phase diagram. Moreover preformed
real-space pairs could coexist with the Fermi-liquid, which
effectively hides them in the normal state kinetics. Any direct
evidence in favor of either scenario is  highly desirable. If
real-space pairs indeed exist in superconducting cuprates, then
their superconducting state should be a three-dimensional (3D)
Bose-Einstein condensate (BEC) of  CBG. Its
 critical behavior \cite{alebook} is rather different from any
"universal" criticality  like   mean-field BCS \cite{bcs}, 3D "XY"
 or KT \cite{kiv} transitions.

Here I show that  high-resolution magnetometery in the critical and
normal regions provides unambiguous evidence for real-space pairing
in cuprates.

 A number of experiments (see, for example,
\cite{mac,jun,hof,nau,igu,ong} and references therein), including
torque magnetometery, showed enhanced diamagnetism above $T_c$.
Originally it was explained as the conventional fluctuation
diamagnetism in quasi-2D BCS superconductors (see, for example Ref.
\cite{hof}). The data taken at relatively low magnetic fields
(typically below 5 Tesla) revealed a crossing point in the
magnetization $M(T,B)$ of most anisotropic cuprates (e.g. Bi-2212),
or in $M(T,B)/B^{1/2}$ of less anisotropic
YBa$_2$Cu$_3$O$_{7-\delta}$ \cite{jun}. The dependence of
magnetization (or $M/B^{1/2}$) on the magnetic field was shown to
vanish at some characteristic temperature below $T_c$. Importantly
more recent data taken in high magnetic fields (up to 30 Tesla) show
that the crossing point, anticipated for low-dimensional
superconductors and associated with conventional superconducting
fluctuations, does not explicitly exist in magnetic fields above 5
Tesla \cite{nau,ong}.

Ref.\cite{ong} has   linked the enhanced normal state diamagnetism
with  mobile vortexes  well above $T_c$ where conventional
fluctuations should be negligible. Surprisingly  the same torque
magnetometery  \cite{mac,nau,ong} uncovered  that the diamagnetic
signal above $T_c$ \emph{increases} in magnitude with applied
magnetic field, $B$. Such magnetic field dependence of magnetisation
$M(T,B)$ is entirely inconsistent  with what one expects from vortex
liquid. While $-M(B)$  decreases logarithmically at temperatures
well below $T_c$, the  experimental curves
 clearly show that $-M(B)$  increases with the
field at and  above $T_c$ , just opposite to what one could expect
in a conventional vortex liquid. These significant departures from
the London liquid behavior  indicates that  vortex liquid does not
appear above the resistive phase transition (see also
Ref.\cite{mac}). Also accepting the vortex scenario and fitting  the
magnetization data in Bi-2212 with the conventional  logarithmic
field dependence \cite{ong}, one obtains surprisingly high upper
critical fields $H_{c2} > 120$ Tesla and a very large
Ginzburg-Landau parameter, $\kappa=\lambda_H/\xi
>450$  even at temperatures close to $T_c$. The in-plane
low-temperature magnetic field penetration depth is
$\lambda_H\approx 220$ nm in optimally doped Bi-2212 (see, for
example \cite{tallon}). Hence the zero temperature coherence length
$\xi$ turns out to be about the lattice constant, $\xi\lesssim
0.5$nm. Such a small coherence length is perfectly compatible with
 direct STM measurements of the individual vortex cores in
Bi-2212 \cite{fis} and  with the size of the vortex core in CBG
\cite{alebook}. However it rules out the "preformed Cooper pairs"
\cite{kiv}, since the pairs are virtually not overlapped at any size
of the Fermi surface.
\begin{figure}
\begin{center}
\includegraphics[angle=-90,width=0.65\textwidth]{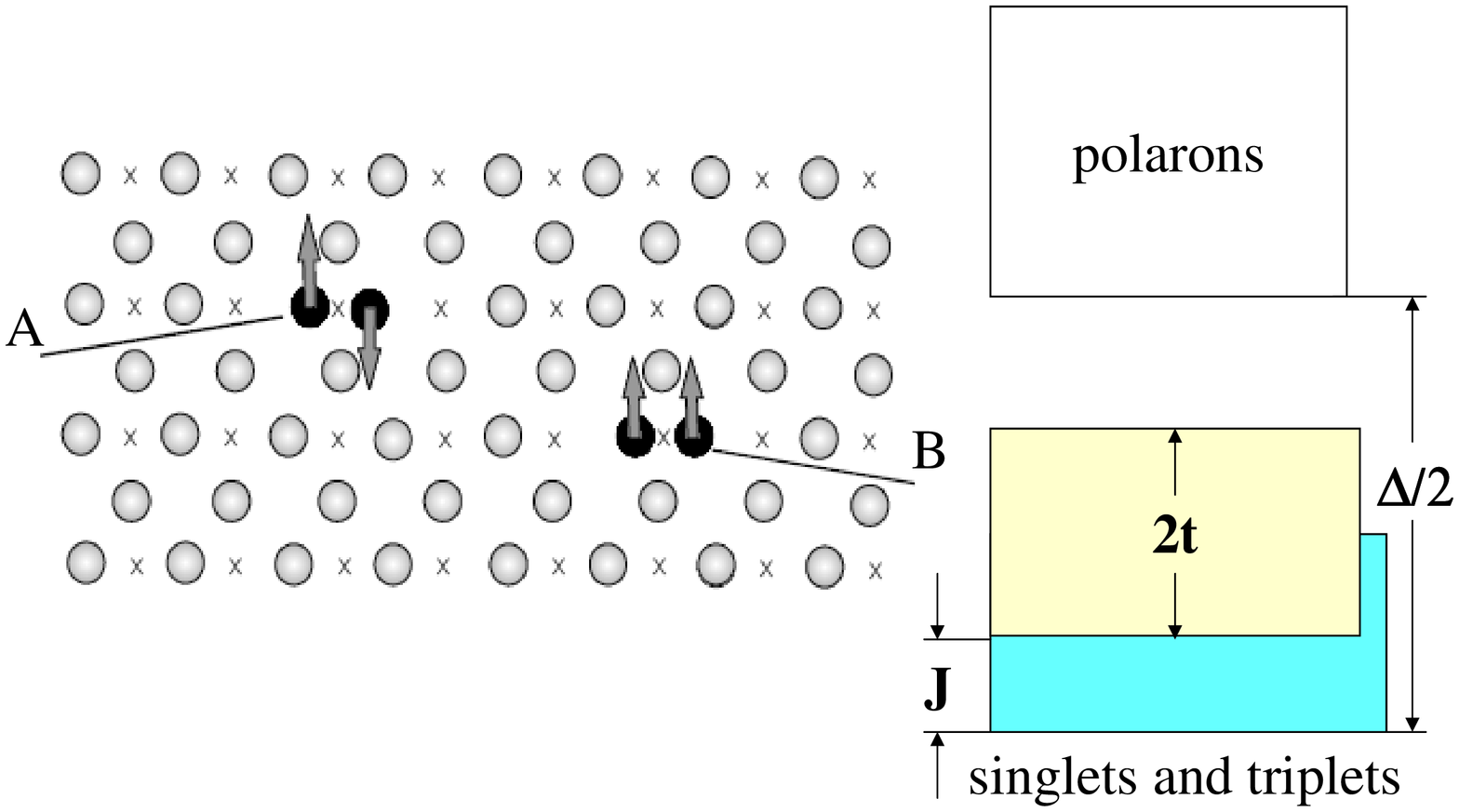}
\vskip -0.5mm \caption{Bipolaron picture of high temperature
superconductors. $A$ corresponds to a singlet oxygen intersite
bipolaron, $B$ is a triplet intersite bipolaron.}
\end{center}
\end{figure}

 Here I calculate the  magnetization, $M(T,B)$, of anisotropic  CBG on a
 lattice,
and compare the result with  diamagnetism of cuprates  recently
measured in Ref. \cite{ong}. A low-energy structure of cuprates in
the bipolaron model
 is shown in Fig.1, where oxygen holes are bound into
real-space intersite singlets (A) and  triplets (B) separated by an
exchange energy $J$ \cite{alemot}, which is estimated as a few tens
or hundreds Kelvin depending on doping  in agreement with
experimental charge and spin pseudogaps in cuprates \cite{kabmic}.
Bipolarons are almost ideal charged bosons, because their Coulomb
repulsion is strongly suppressed by a large lattice dielectric
constant.
 When the
 magnetic field is applied perpendicular to  copper-oxygen
plains the quasi-2D bipolaron energy spectrum is quantized as
\begin{equation}
E_\alpha= \omega(n+1/2) +2t_c [1-\cos(K_zd)],
\end{equation}
where $\alpha$ comprises $n=0,1,2,...$ and  in-plane $K_x$ and
out-of-plane $K_z$ center-of-mass quasi-momenta, $\omega=2\hbar
eB/\sqrt{m_xm_y}$, $t_c$ and $d$ are the hopping integral and the
lattice period perpendicular to the planes. The spectrum consists of
two degenerate brunches, the so-called $"x"$ and $"y"$ bipolarons
\cite{alebook}, with anisotropic in-plane bipolaron masses
$m_x\equiv m$ and $m_y\approx 4m$. Expanding the Bose-Einstein
distribution function in powers of $exp[(\mu-E)/k_BT]$ with the
negative chemical potential $\mu$ one can after summation over $n$
readily obtain
 the boson density
\begin{equation}
n_b={2eB\over{\pi \hbar d}} \sum_{r=1}^{\infty} I_0(2t_c r/k_BT)
{\exp[ (\tilde{\mu} -2t_c)r/k_BT]\over{1-\exp(-\omega r/k_BT)}},
\end{equation}
and the magnetization, $M(T,B)= -k_BT\partial/\partial B
\sum_{\alpha} \ln[1-\exp(\mu-E_\alpha)/k_BT]$,
\begin{eqnarray}
&&M(T,B)=-n_b \mu_b+  {2ek_BT\over{\pi \hbar d}} \sum_{r=1}^{\infty}
I_0\left({2t_c r\over {k_BT}}\right)\times \\
&&{\exp[ (\tilde{\mu} -2t_c)r/k_BT]\over{1-\exp(- \omega r/k_BT)}}
\left({1\over{r}}-{\omega \exp(-\omega
r/k_BT)\over{k_BT[1-\exp(-\omega r/k_BT)]}}\right),\nonumber
\end{eqnarray}
where $\mu_b=\hbar e/\sqrt{m_xm_y}$, $\tilde{\mu}=\mu-\omega/2$ and
$I_0(x)$ is the modified Bessel function. At low temperatures $T
\rightarrow 0$ Schafroth's result \cite{sha} is recovered, $M(0,B)=
-n_b \mu_b$. The magnetization of charged bosons is
field-independent at low temperatures. At high temperatures, $T \gg
T_c$ the chemical potential has a large  magnitude, and we can keep
only the terms with $r=1$ in Eqs.(2,3) to obtain
\begin{figure}
\begin{center}
\includegraphics[angle=-90,width=0.65\textwidth]{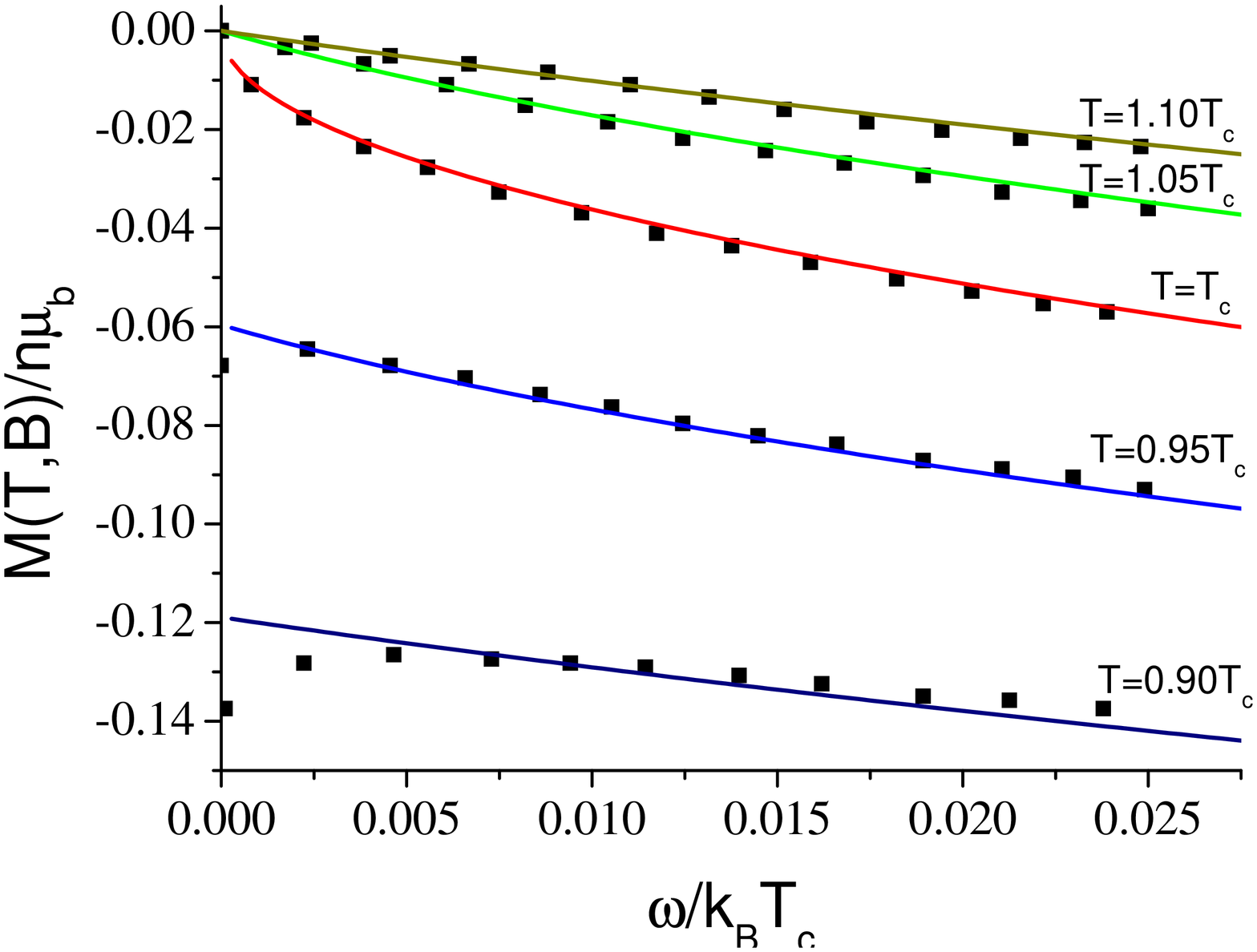}
\vskip -0.5mm \caption{Exact numerical magnetization (symbols)
compared with the analytical approximation, Eq.(4) (lines) .}
\end{center}
\end{figure}
   $M(T,B)=-n_b \mu_b \omega/(6k_BT)$ at $k_BT \gg k_BT_c\gg \omega$,
 which is the familiar  Landau  orbital diamagnetism  of nondegenerate
 carriers.

The critical region $\tau=T/T_c-1\ll 1$  requires numerical
calculations, which have been done \cite{aleden} for an anisotropic
3D CBG with $t_c\gtrsim k_BT_c/2$ and $I_0(x)\approx e^x/\sqrt{2\pi
x}$ in Eqs.(2,3). Notwithstanding, one can nicely map the exact
results, Fig.2, with a simple analytical expression by replacing
summation
 over all but the first Landau level for integration,
\begin {equation}
{M(T,B)\over{n_b
\mu_b}}=-{0.22\omega\over{k_BT_c}}\left[\tau+\sqrt{{0.37
\omega\over{k_BT_c}} +\tau^2}\right]^{-1}.
\end{equation}
Remarkably Eq.(4) predicts almost field-independent diamagnetism
well below $T_c$, $|\tau|\gg \omega/k_BT_c$, a linear field
dependence $M(T_c,B) \sim B$ well above $T_c$, $\tau\gg
\omega/k_BT_c$, and an unusual square root behavior at $T=T_c$,
$M(T_c,B) \sim B^{1/2}$. Here $T_c$ is the familiar Bose-Einstein
condensation temperature $k_{B}T_{c}= 3.31\hbar
^{2}(n_{b}/2)^{2/3}/(m_{x}m_{y}m_{c})^{1/3}$, with $m_{c}=\hbar
^{2}/|t_{c}|d^{2}$.

\begin{figure}
\begin{center}
\includegraphics[angle=-90,width=0.65\textwidth]{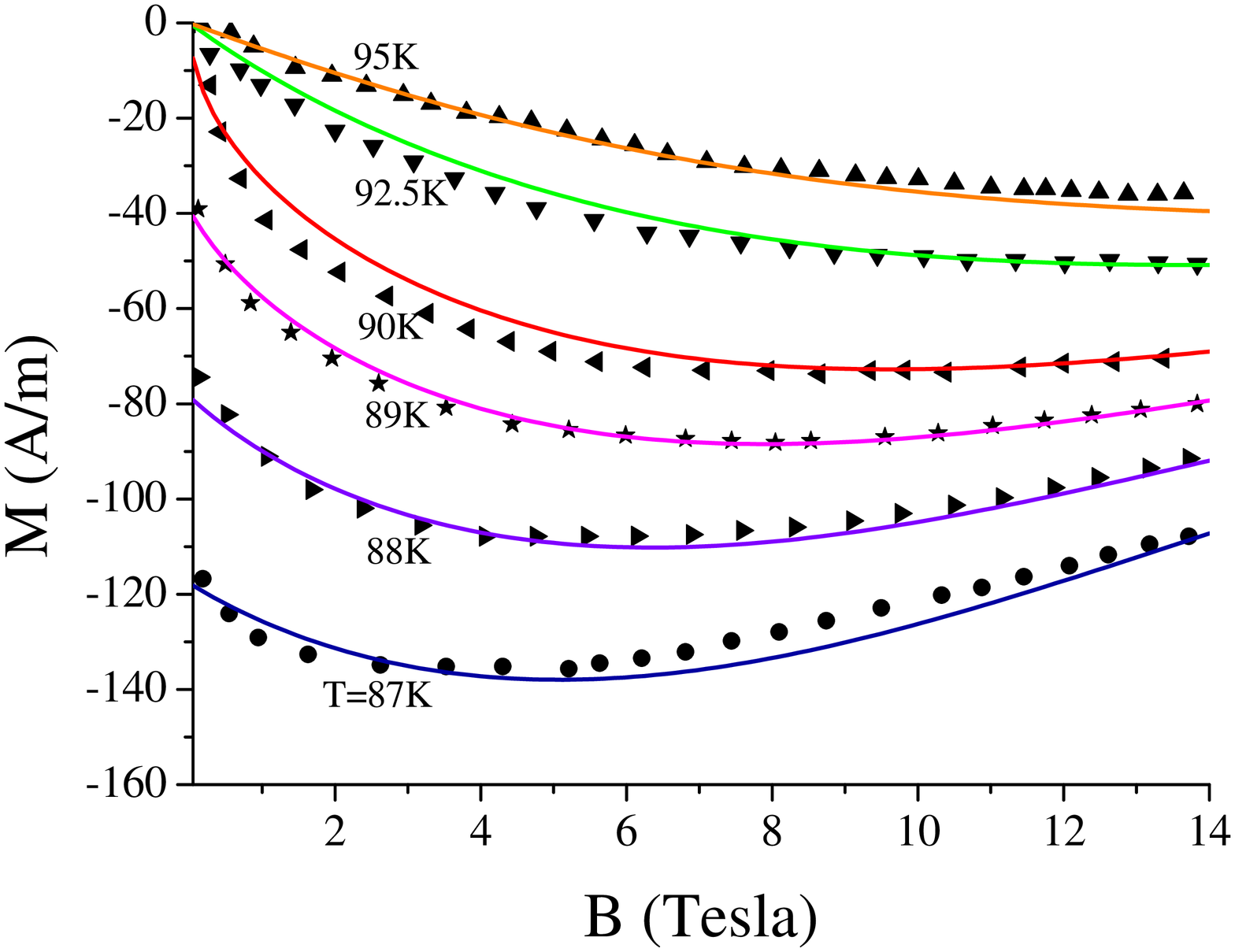}
\vskip -0.5mm \caption{Diamagnetism of optimally doped Bi-2212
(symbols)\cite{ong} compared with magnetization of CBG, Eq.(6), near
and above $T_c$ (lines).}
\end{center}
\end{figure}
Comparing with experimental data one has to take into account a
temperature and field depletion of singlets due to their thermal
excitations into spin-split triplets and single polaron states,
Fig.1B. If the spin gap is small compared with the charge pseudogap,
$J <\Delta/2$,
 triplets mainly contribute to  temperature and field dependencies
of the singlet bipolaron density near $T_c$,
 \begin{equation}
 n_b(T,B)=n_b(T_c,0)[1-\alpha \tau -(B/B^*)^2],
 \end{equation}
 where
 $\alpha=3(2n_ct)^{-1}[J (e^{J/k_BT_c}-1)^{-1}-k_BT_c\ln(1-e^{-J/k_BT_c})]$,
 $\mu_BB^*=(2k_BT_cn_ct)^{1/2}
\sinh(J/2k_BT_c)$, $\mu_B\approx 0.93 \times 10^{-23}$ Am$^2$ is the
 Bohr magneton, $n_c$ is the atomic density of singlets at $T=T_c$  in zero
 field ($n_c \lesssim 0.1$ in optimally doped cuprates),
 and $2t$ is the triplet bandwidth, which is taken
 much larger than $k_BT_c$. A triplet contribution to diamagnetism remains negligible compared with the singlet
 diamagnetism if $\tau \ll J/k_BT_c$. Then  Eq.(4)
 mapping
  numerical  magnetization in the  critical region is  modified as
\begin{eqnarray}
&&M(T,B)=-{0.22n_b(T_c,0) \mu_b  B\over {B_0(1+2\alpha/3)}}\times  \\
&&\left[\tau +{(B/B^*)^2\over{1+2\alpha/3}}+\sqrt{{0.37
B/B_0\over{(1+2\alpha/3)^2}} +\left(\tau +{(B/B^*)^2\over
{1+2\alpha/3}}\right)^2}\right]^{-1}, \nonumber
\end{eqnarray}
where $B_0=k_BT_c/2\mu_b$. Using the magnetic field in-plane
penetration depth, $\lambda_H^{-2} \approx 21$ $(\mu m)^{-2}$ of
optimally doped Bi-2212 \cite{tallon} and of CBG \cite{alebook},
$\lambda_H^{-2}=2n_be^2(m_x+m_y)/(\mu_0m_xm_y)$, we  estimate the
bipolaron mass as $m\approx 7.5 m_e$ in agreement with the
analytical and numerical QMC results \cite{alebook},  and
$n_b(T_c,0) \mu_b=\hbar \mu_0 (m_xm_y)^{1/2}/2e \lambda_H^2
(m_x+m_y) \approx 2100$A/m. The BEC temperature corresponds to the
temperature were the in-plane resistivity starts to drop with
temperature lowering, which   is about $T_c=90$K in optimally doped
Bi-2212, so that $B_0=524$ Tesla. This choice of $T_c=90$K is also
justified by low-field magnetization \cite{ong}, which has an
exponent close to 1/2, $M(90K,B)\sim B^{1/2}$ just for this
temperature. The remaining two parameters in Eq.(6) are found using
the experimental  field dependence of $M(T,B)$ at any fixed
temperature near $T_c$. Fitting $M(T,B)$ at $T=89$ K, Fig.3, yields
$\alpha=0.62$ and $B^*=56$ Tesla, which according to Eq.(5)
corresponds to the singlet-triplet exchange energy $J\approx 20$K.
 Quite remarkably all other experimental curves in the critical
region  are well described by Eq.(6) without any fitting parameters,
Fig.3.

I conclude that the normal state diamagnetism observed in many
cuprates \cite{mac,jun,hof,nau,igu,ong} provides unambiguous
evidence for charged real-space bosons. The experimental data,
Fig.3, clearly contradict   BCS (with or without conventional
fluctuations) and KT scenarios of the phase transition in cuprates.
If we define a critical exponent as $\delta=\ln B/\ln|M(T,B)|$ for
$B\rightarrow 0$, the $T$ dependence of $\delta(T)$ in CBG is
dramatically different from KT and other "universal" critical
exponents, but it is very close to the experimental \cite{ong}
$\delta(T)$, Fig.4.
\begin{figure}
\begin{center}
\includegraphics[angle=-90,width=0.55\textwidth]{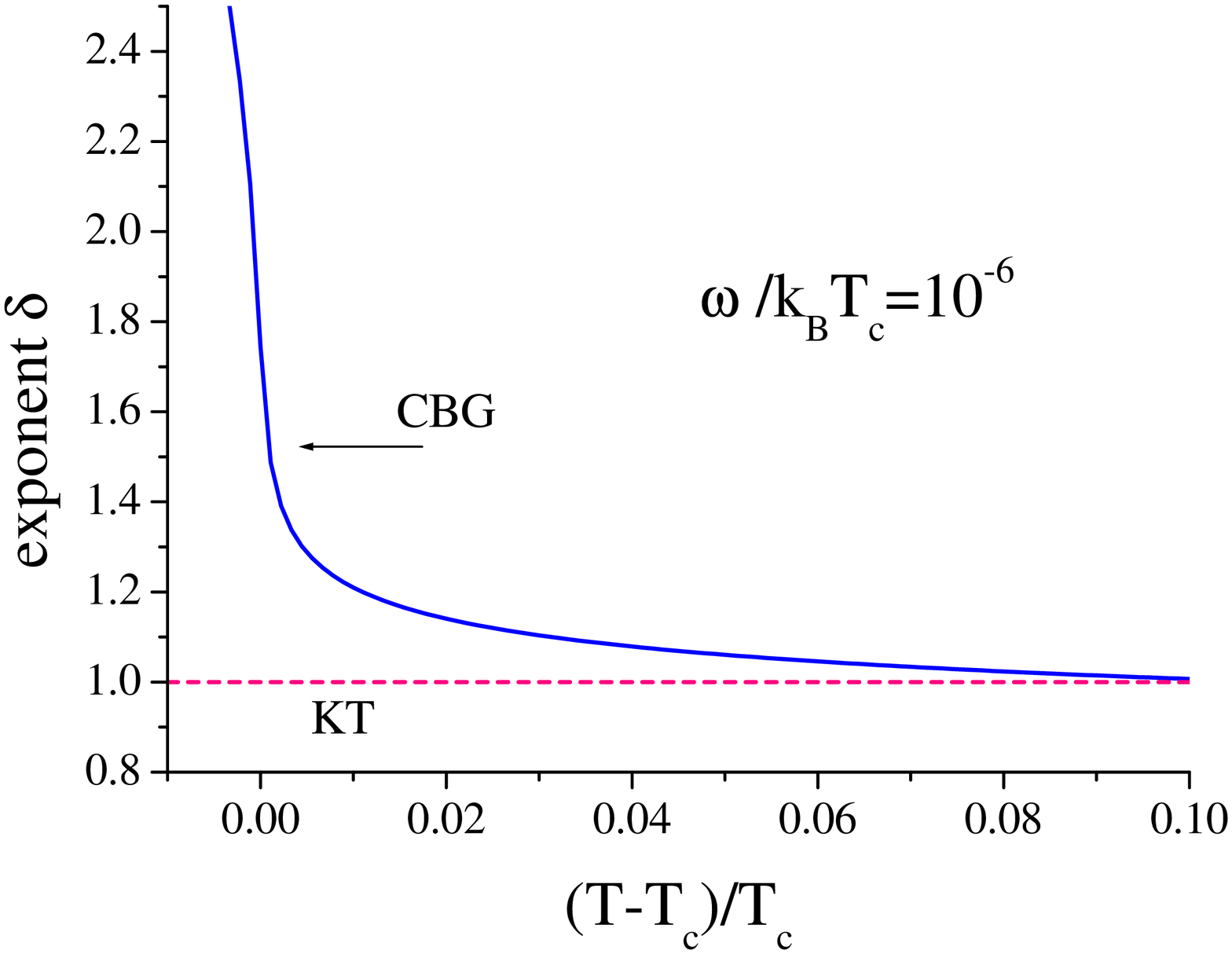}
\vskip -0.5mm \caption{Critical exponents of the low-field
magnetization in CBG and in KT transition.}
\end{center}
\end{figure}

Another strong argument in favor of 3D BEC in cuprates has been
drawn using parameter-free fitting of experimental $T_c$ with BEC
$T_{c}$ in more than 30 underdoped, optimally and overdoped samples
\cite{alekab}. Whereas the  KT critical temperature expressed
through the in-plane penetration depth \cite{uem,kiv}
$k_{B}T_{KT}\approx 0.9d\hbar ^{2}/(16\pi e^{2}\lambda _{H}^{2})$
appears  several times higher than the experimental
values in many cases. There are also quite a few samples with about the same $%
{\lambda _{H}}$ and the same $d$, but with very different values of
$T_{c},$ in disagreement with  the KT transition. The large Nernst
signal, allegedly supporting vortex liquid in the normal state of
cuprates \cite{xu}, has been  explained as perfectly normal state
phenomenon owing to a partial localization of charge carriers in a
random potential inevitable in cuprates \cite{alezav}.  CBG upper
critical field  and the specific heat in the magnetic field have
been found in striking consensus with experimental data \cite{ZAV}
following our prediction \cite{aleH}. More recently the d-wave
symmetry and real-space modulations of  the order parameter have
been also explained with CBG  in underdoped \cite{alebook} and
overdoped \cite{and} cuprates.

I thank  Victor Kabanov  for his help with numerical calculations
  and valuable discussions. The work was supported by  EPSRC (UK) (grant
EP/C518365/1).


\begin{thebibliography}{90}
\bibitem{bed}  J.G. Bednorz and K.A. M\"{u}ller, Z. Phys. B%
{\bf 64}, 189 (1986).
\bibitem{ogg} R.A. Ogg Jr., Phys. Rev. {\bf 69}, 243
(1946).
\bibitem{sha} M.R. Schafroth, Phys. Rev. {\bf 100}, 463 (1955); J.M.  Blatt and S.T. Butler,  Phys. Rev. {\bf %
100}, 476 (1955).
\bibitem{bcs} J. Bardeen, L.N. Cooper, and J.R. Schrieffer,
Phys. Rev {\bf 108}, 1175 (1957).
\bibitem{aleran}
A. Alexandrov  and J. Ranninger, Phys. Rev. B{\bf 23}, 1796 (1981),
ibid {\bf 24}, 1164 (1981).
\bibitem{ale} A.S. Alexandrov, Russ. J. Phys. Chem. {\bf 57}, 167
(1983).
\bibitem{zhao}  G.M. Zhao and D.E. Morris
Phys. Rev. B {\bf 51}, R16487 (1995); G.-M. Zhao, M. B. Hunt, H.
Keller, and K. A. M\"uller, Nature (London) {\bf
 385}, 236 (1997); R. Khasanov, D. G. Eshchenko, H. Luetkens, E. Morenzoni, T. Prokscha, A. Suter,
 N. Garifianov, M. Mali, J. Roos, K. Conder, and H. Keller
Phys. Rev. Lett. {\bf 92}, 057602 (2004).
\bibitem{LAN}  A. Lanzara, P.V. Bogdanov, X.J. Zhou, S.A.
Kellar, D.L. Feng, E.D. Lu, T. Yoshida, H. Eisaki, A. Fujimori,K.
Kishio, J.I. Shimoyana, T. Noda, S. Uchida, Z. Hussain and Z.X.
Shen, Nature (London) {\bf 412}, 510 (2001).
\bibitem{mic}  D. Mihailovic, C.M. Foster, K. Voss, and A.J. Heeger, Phys. Rev. B{\bf 42}, 7989 (1990).
\bibitem{ega} T.R. Sendyka, W. Dmowski,  T. Egami, N. Seiji, H. Yamauchi, and S. Tanaka, Phys. Rev. B{\bf 51}, 6747
(1995).
\bibitem{alebook} A.S. Alexandrov,
\emph{Theory of Superconductivity: From Weak to Strong Coupling}
(IoP Publishing, Bristol and Philadelphia, 2003), and references
therein.
\bibitem{kiv}  V.J. Emery and S.A. Kivelson, Nature (London), {\bf
374}, 434 (1995).
\bibitem{mac} C. Bergemann, A.W. Tyler, A.P. Mackenzie,
J. R. Cooper, S.R. Julian, and D.E. Farrell, Phys. Rev. B {\bf 57},
14387 (1998).
\bibitem{jun} A. Junod, J-Y. Genouda, G. Trisconea, and T. Schneider,   Physica C {\bf 294}, 115 (1998).
\bibitem{nau} M. J. Naughton, Phys. Rev. B{\bf 61}, 1605 (2000).
\bibitem{hof} J. Hofer, T. Schneider, J.M. Singer, M. Willemin, H. Keller,
 T. Sasagawa, K. Kishio, K. Conder, and J. Karpinski, Phys. Rev. B {\bf 62},  631 (2000).
\bibitem{igu} I. Iguchi, A. Sugimoto, and H. Sato, J. Low Temp. Phys. {\bf 131}, 451 (2003).
\bibitem{ong} Y. Wang, L. Li, M.J. Naughton, G.D. Gu, S. Uchida, and
N.P. Ong, cond-mat/0503190; L. Li, Y. Wang, M.J. Naughton, S. Ono,
Y. Ando, and N.P. Ong, cond-mat/0507617; Y. Wang, L. Li and N.P.
Ong, cond-mat/0510470.
\bibitem{fis} B.W. Hoogenboom, M. Kugler, B. Revaz, I. Maggio-Aprile, O. Fischer, and Ch.
Renner, Phys. Rev. B {\bf 62}, 9179 (2000).
\bibitem{alemot}  A.S. Alexandrov and N.F. Mott, J. Supercond (US), {\bf 7},
599 (1994).
\bibitem{kabmic} D. Mihailovic, V.V. Kabanov, K. Zagar,
and J. Demsar, Phys. Rev. B{\bf 60}, R6995 (1999) and references
therein.
\bibitem{aleden} C.J.
Dent, A.S. Alexandrov, and V.V. Kabanov, Physica C{\bf 341-348}, 153
(2000).
\bibitem{tallon} J.L. Tallon, J.W. Loram, J.R. Cooper, C. Panagopoulos, and C.
Bernhard, Phys. Rev. B {\bf 68}, 180501(R) (2003).
\bibitem{alekab} A.S. Alexandrov and V.V. Kabanov
Phys. Rev. B {\bf 59}, 13628 (1999).
\bibitem{uem} Y.J. Uemura et al.,
Phys. Rev. Lett. {\bf 62}, 2317 (1989).
\bibitem{xu} Z.A. Xu, N.P. Ong, Y. Wang, T. Kakeshita, and S.
Uchida, Nature (London) {\bf 406}, 486 (2000).
\bibitem{alezav}
A.S. Alexandrov and V.N. Zavaritsky, Phys. Rev. Lett. {\bf 93},
217002 (2004).
\bibitem{ZAV}  V.N. Zavaritsky, V.V. Kabanov and A.S.
Alexandrov, Europhys. Lett. {\bf 60}, 127 (2002).
\bibitem{aleH} A.S. Alexandrov,  Phys. Rev. B{\bf 48},
10571 (1993) and references therein.
\bibitem{and} A.F. Andreev, Pis'ma Zh. Eksp. Teor. Fiz. {\bf 79}, 100 (2004).

\end{thebibliography}
\end{document}